\documentclass{article}
\usepackage{graphicx}
\begin{document}

\def\be{\begin{equation}}
\def\ee{\end{equation}}
\def\ba{\begin{eqnarray}}
\def\ea{\end{eqnarray}}

\title{Statistical fluctuations of ground--state energies and binding
  energies in nuclei}

\author{A. Molinari and H. A. Weidenm\"uller}

\maketitle
\begin{abstract}
The statistical fluctuations of the ground--state energy and of the
binding energy of nuclei are investigated using both perturbation
theory and supersymmetry. The fluctuations are induced by the
experimentally observed stochastic behavior of levels in the vicinity
of neutron threshold. The results are compared with a recent analysis
of binding--energy fluctuations by Bohigas and Leboeuf, and with
theoretical work by Feshbach {\it et al}.  
\end{abstract}

{\it Introduction.} In this letter, we address the question: How does
the interaction with high--lying configurations affect the properties of
nuclear ground states? In particular, what are the ensuing uncertainties,
given the fact that not much is known about both, the interaction and the
high--lying configurations? The question has arisen within two seemingly
different contexts. (i) In a purely theoretical framework, and inspired
by the analogy with the theory of the optical model (where averaging
over the random fluctuations of the compound--nucleus resonances is an
essential element), the question was addressed using a statistical model
for the high--lying states. To this end, Feshbach's projection operator
method~\cite{fes} has been extended to the bound--state problem and used
to estimate the resulting uncertainties in ground--state energies and
wave functions~\cite{car}. (ii) A series of ever--refined nuclear mass
formulas with a sizeable number of parameters has been used to fit the
known binding energies. In spite of many years of dedicated effort,
there remains a difference between mass formulas and data: The data
fluctuate about the (smooth) best fits. The fluctuations have a width
of about $0.5$ MeV and have been interpreted as being due to chaotic
nuclear motion~\cite{boh}. Since the spectral fluctuations of nuclear
levels near neutron threshold (and near the Coulomb barrier for protons)
follow random--matrix predictions~\cite{boh1}, the question arises
whether this fact can be used to account for the observed fluctuations.

We present a novel statistical approach to the question formulated
above. In contrast to Refs.~\cite{car}, it is not our aim to estimate
the theoretical uncertainty of nuclear binding energies resulting from
the use of a mean--field approach. Rather, we are interested in the
statistical fluctuations of ground--state and binding energies which
are induced by those of high--lying configurations. These represent
genuine theoretical uncertainties which are quite independent of the
particular approach used to calculate nuclear properties. To this end
we describe the high--lying configurations in terms of a random--matrix
model. In trying to be as conservative as possible, we only use the
fact just mentioned that in many nuclei with mass number $A > 30$ or
so, random--matrix theory correctly describes the statistical
fluctuations of states at excitation energies of 8 to 10 MeV. We show
that from this input it is possible to estimate the minimum
statistical uncertainties in nuclear ground--state energies and, from
there, nuclear binding energies. We compare our results with the work
of Refs.~\cite{car}, and with the empirical results of Ref.~\cite{boh}.

\begin{figure}
\begin{center}
\includegraphics[clip,height=3cm]{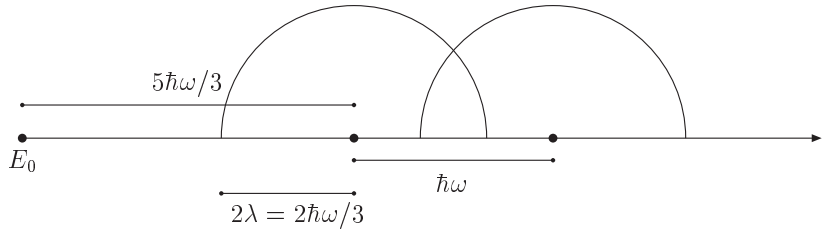}
\caption{\label{fig1}\textit{Schematic representation of
    the nuclear spectrum in the shell model. The semicircles represent
 the shells with $n = 1$ and $n = 2$.} 
 }
\end{center}
\end{figure}

{\it Model.} Starting point is the nuclear shell model. Neighboring
major shells are separated by an energy difference $\hbar \omega
\approx 41 / A^{1/3}$ MeV. We first neglect the mixing between major
shells due to the residual interaction. Then, the nuclear ground state
is obtained by diagonalizing the shell--model Hamiltonian (which
consists of the single--particle energies of the lowest shell and of
the matrix elements of the residual interaction with respect to
many--body states in that shell). The resulting ground state has
energy $E_0$. The corresponding eigenfunction is composed of
configurations belonging to the lowest major shell. We assume that the
residual interaction within the states of each of the remaining major
shells (excitation energies $n \hbar \omega$ with $n = 1, 2, \ldots$)
is so strong as to completely mix the configurations within that shell.
Then, the spectra and eigenfunctions within each such shell are
described by random--matrix theory, more precisely: by the Gaussian
orthogonal ensemble (GOE) of random matrices. For $n \geq 1$, this
assumption is consistent with the spectral fluctuations analyzed in
Ref.~\cite{boh1}, and with the properties of nuclear cross--section
fluctuations at higher excitation energies~\cite{eri}. The resulting
nuclear spectrum is schematically depicted in
Fig.~\ref{fig1}.
Each of the higher shells ($n \geq 1$) gives rise to a Gaussian--shaped
mean level density. In what follows we replace each Gaussian by the
semicircle characteristic of random--matrix theory. For our
order--of--magnitude estimates, the difference should be irrelevant. We
focus attention on the shell with $n = 1$ and later show that higher
shells do not affect the result significantly. We estimate the shift
and the uncertainty of the ground--state energy $E_0$ using (i)
perturbation theory, (ii) the supersymmetry approach. Here, $E_0 < 0$
is the difference between the location of the ground state and the
center of the shell with $n = 1$. The semicircle has radius $2 \lambda$.
Estimates for $E_0$ and of $\lambda$ are given below. For simplicity,
we focus attention entirely on the ground state and omit the other
states in the shell with $n = 0$.

{\it Perturbation Theory.} We denote by $V_{0 j}$ the matrix elements
of the residual interaction connecting the ground state with the
states $j= 1, \ldots, N$ in the major shell with $n = 1$, and by $E_j$
the GOE eigenvalues in that shell, counted from the center. For $N$,
the number of states in that shell, we have $N \gg 1$. The energy
$\tilde{E_0}$ of the perturbed ground state is given by
\be
\tilde{E_0} = E_0 + \sum_{j = 1}^N \frac{V^2_{0 j}}{E_0 - E_j} + \ldots\
\ . 
\label{1}
\ee
Within the GOE, the $V_{0 j}$'s and the $E_j$'s are uncorrelated random
variables. The $V_{0 j}$'s are Gaussian distributed with zero mean
value. Using the standard definition $\Gamma^{\downarrow} = 2 \pi
\overline{V^2_{0 j}} / d$ of the spreading width for mixing of the
ground--state wave function with the states of the $n = 1$ shell, we
write the second moment $\overline{V_{0 j}V_{0 k}} = (1/(2 \pi))
\Gamma^{\downarrow} d \ \delta_{j k}$. Here, $d = \pi \lambda / N$ is
the mean level spacing in the $n = 1$ shell. Averaging Eq.~(\ref{1})
over the GOE, we find to lowest order in $\Gamma^{\downarrow}$,
\be
\overline{\tilde{E_0}} \approx E_0 + \frac{\Gamma^{\downarrow} d}{2
\pi} \int {\rm d} E^{\prime} \frac{1}{E_0 - E^{\prime}}
\overline{\sum_j \delta(E^\prime - E_j)} \ .
\label{2}
\ee
By definition, $\overline{\sum_j \delta(E^\prime - E_j)}$ is equal to
$\overline{\rho(E^\prime)}$, the average level density normalized to
$N$. Hence we find approximately
\be
\overline{\tilde{E_0}} \approx E_0 + \frac{\Gamma^{\downarrow}
\lambda}{2 E_0} \ .  
\label{3}
\ee
The level shift $\Gamma^{\downarrow} \lambda / (2 E_0)$ is negative,
as expected. To calculate the variance, we define
$\overline{\rho(E_1,E_2)} = \sum_{j k} \overline{\delta(E_1 - E_j)
\delta(E_2 - E_k)}$ and use the fact that the correlation function
$\overline{\rho(E_1,E_2)} - \overline{\rho(E_1)} \ \overline{\rho(E_2)}$
tends to zero over distances $|E_1 - E_2|$ that are of order $d$ (and
not $N d$). Proceeding as before, we find
\be
\overline{(\tilde{E_0} - E_0)^2} - \biggl(\overline{\tilde{E_0}} -
E_0\biggr)^2 = \frac{c}{N} \biggl( \frac{\Gamma^{\downarrow}
\lambda}{2 E_0} \biggr)^2 \ .
\label{4}
\ee
Here $c$ denotes a constant of order unity (and not of order $N$). 
We observe that with $N \gg 1$, the fluctuations are much smaller
than the level shift. Rough estimates are obtained by putting $2
\lambda \approx (2/3) \hbar \omega$ (the semicircles of the shells
$n$ and $n + 1$ overlap), $E_0 \approx - (5/3) \hbar \omega$, and
$\Gamma^{\downarrow} \approx 3$ MeV. The last figure roughly
corresponds to the spreading width of the giant dipole resonance
which accounts for the mixing of one--particle one--hole and
two--particle two--hole configurations and is, thus, similar to the
mixing of the ground--state wave function with the wave functions in
the shell with $n = 1$ considered here. For the shift, this yields a
typical value of $- 0.3$ MeV. The r.m.s. fluctuation is roughly given
by that value divided by $\sqrt{N}$. To estimate $N$ we use $N = \pi
\lambda / d_t$. Here $d_t$ is the mean level spacing at neutron
threshold. For heavy nuclei, $d_t \approx 10$ eV, $\hbar \omega
\approx 6.5$ MeV, and $N \approx 10^6$ while for nuclei with mass
number $A$ around 50 we have $d_t \approx 10$ keV, $\hbar \omega
\approx 10$ MeV and $N \approx 10^3$. We find that the r.m.s.
fluctuation is about 0.3 keV for heavy nuclei and about 10 keV for
nuclei with $A \approx 50$. These figures are order--of--magnitude
estimates only, of course.

{\it Supersymmetry.} We use the supersymmetry method
(SUSY)\cite{efe,ver} in order to show that the perturbative estimates
given above essentially remain valid beyond the domain of
perturbation theory. The calculations are rather technical and cannot
be given here. A more detailed account is in preparation~\cite{alf}.
Suffice it to say that using SUSY and $N \gg 1$, we find the following
saddle--point equation for the $\sigma$ matrix of the one--point
function:
\be
\sigma = \frac{\lambda}{E - \lambda \sigma} + \frac{1}{N} \
\frac{\lambda}{E - \lambda \sigma - (1/2) \Gamma^{\downarrow}
\lambda / (E - E_0)} \ .  
\label{5}
\ee
Here the first term on the r.h.s. is the usual contribution which,
taken by itself, results in the semicircle law. (We recall that
the spectrum is located in that energy interval where $\sigma$ has
non--real values, with the average level density proportional to
${\rm Im} \ \sigma$). The second term is the correction due to the
coupling with the ground state. This term yields an additional
contribution to ${\rm Im} \ \sigma$ in an energy interval of width
$\Delta E$ located at $\tilde{E_0}$ outside the semicircle and in
the vicinity of $E_0$. We solve Eq.~(\ref{5}) for the diagonal
elements of the $\sigma$ matrix using perturbation theory. For the
shift, we retrieve our previous result, Eq.~(\ref{3}). For
$\Delta E$ we find
\be
\Delta E = \frac{1}{N^{1/2}} \ \frac{\Gamma^{\downarrow}
\lambda}{2 |E_0|} \ \frac{4 \lambda}{|E_0|} \ .
\label{6}
\ee
This agrees with Eq.~(\ref{4}) except for the last factor. This
factor is of order unity. It arises because in Eqs.~(\ref{4}) and
(\ref{6}), we have calculated different expressions (the r.m.s.
fluctuation and the width of the spectrum, respectively). We have
used Eq.~(\ref{5}) also to calculate shift and width
non--perturbatively for values of $E_0$ close to the edge of the
semicircle. We have found that these values differ from the
perturbative ones only by numerical factors. We conclude that our
estimates in Eqs.~(\ref{3}) and (\ref{5}) are approximately valid
in a wide range of values for $E_0$ and $\Gamma^{\downarrow}$.

{\it Discussion.} So far, only the influence of the shell with
$n = 1$ on the ground--state energy has been taken into account.
What about higher shells ($n > 1$)? Since SUSY has corroborated the
perturbative results, we may safely use perturbation theory to
answer the question. Formally, Eqs.~(\ref{3}) and (\ref{4}) remain
valid, with $E_0$ replaced by the distance $D_n$ between the energy
of the ground state and the center of the shell $n$ under
consideration, and $\Gamma^{\downarrow}$ replaced by
$\Gamma^{\downarrow}_n$, the spreading width describing the mixing
of the ground--state wave function with the configurations in shell
$n$. While $(D_n)^{-1}$ decreases only like $n^{-1}$,
$\Gamma^{\downarrow}_n$ falls off very rapidly (exponentially).
This is because $\Gamma^{\downarrow}$ depends partly upon the
overlap between the ground--state wave function and the
configurations in shell $n$. In a different context, this exponential
fall--off has been demonstrated numerically in Figure~1 of
Ref.~\cite{zel} where the dependence of the ground--state energy upon
the truncation of the shell--model space was studied. Therefore, we
expect that the inclusion of higher shells would affect our estimates
only by a factor two or three or so. It is possible, of course, that
our treatment is too conservative, and that the domain where the GOE
applies, extends further down in the spectrum. This would reduce
$|E_0|$ and $N$ and, thus, enhance both the shift and the r.m.s.
fluctuation. Unfortunately, there is hardly any empirical evidence
one way or another concerning that problem. In summary, we find that
the shift of the ground--state energy is of the order of or larger
than 0.5 to 1 Mev for all nuclei, while the fluctuations amount at
least to several 10 keV (1 keV) for medium--weight (heavy) nuclei,
respectively. It goes without saying that these fluctuations do not
describe actual fluctuations of the ground--state energy, but rather
the minimum theoretical uncertainty due to lack of precise
information about the states and interactions in the shells with $n
\geq 1$.

So far we have considered only a single level rather than the
complete set of states belonging to the lowest shell. We do not
expect that the presence of these states strongly affects our
estimates. To be sure that this is indeed the case, we should also
derive and solve the saddle--point equation for this case. This work
is now under way~\cite{alf}.

How are our results related to those of Refs.~\cite{car}? The problem
addressed in Refs.~\cite{car} was not to compute the r.m.s.
fluctuations, but rather to relate the theoretical uncertainty of the
binding energy per nucleon to that of the mean--field energy and to
that of the residual interaction (the difference between the true
interaction and the mean field). The latter connects the mean--field
ground state to the excited states. This was done in the framework of
a somewhat simpler problem, the energy per particle of nuclear matter,
rather than for finite nuclei. 

In most nuclei, the mean--field energy differs from the energy denoted
by $E_0$ in the present paper. Indeed, our definition of $E_0$
includes the effects of the residual interaction within the lowest
shell and differs, therefore, from the mean--field energy. Moreover,
we focus attention on stochastic uncertainties resulting from the
chaotic behavior of excitet states in the vicinity of neutron
threshold. We do not attempt to estimate the error that comes with a
theory based upon a mean--field approach. Therefore, the approach of
Refs.~\cite{car} and the one used here, are rather different, and a
direct comparison of the results is not possible.

{\it Fluctuations of Binding Energies.} How can we use our results to
estimate the fluctuations in the binding energies of nuclei? The
binding energy is the minimum excitation energy at which the nucleus
fragments into $A$ separate nucleons. We do not believe that it is
possible at present to determine this excitation energy reliably in
the framework of a nuclear--structure calculation for nucleus $A$.
Indeed, to the best of our knowledge, nuclear--structure calculations
have rarely been used to calculate binding energies, except for the
lightest nuclei. To answer the questions raised in Ref.~\cite{boh}, we
use the following indirect approach. Our calculations yield the
fluctuation of the nuclear ground--state energy with respect to the
center of the shell with $n = 1$. For most nuclei, the threshold for
neutron or for proton emission lies somewhere in that shell. Thus, we
identify the calculated fluctuation with the fluctuation of the
ground--state energy with respect to neutron (or proton) threshold.
The fluctuation of the nuclear binding energy is then obtained by the
combined effect of removing successively $A$ nucleons from the nucleus.

We apply this scheme in reverse order, starting from a nucleus with
mass number $A_0$ and adding $A - A_0$ nucleons to it. The first
nucleus (in increasing order of $A$) where we have definitive
evidence (both experimentally and theoretically) that GOE statistics
applies is $^{26}$Al~\cite{Mitchell,Zelevinsky}. We assume that
nuclei with mass numbers $A \leq 25$ do not have any fluctuations
in either ground--state or binding energy. Therefore, we choose $A_0
= 26$ and an initial value 30 keV for the fluctuation of the
ground--state energy and, thus, of the binding energy. To calculate the
fluctuations of the binding energy for nuclei with mass numbers $A >
A_0$ we assume, for simplicity, that the fluctuations of the
ground--state energy have a Gaussian distribution. (In the framework
of our model, this may or may not be the case. We have not explored
the solutions of the saddle--point equation to the point of
ascertaining the answer to this question but recall that for the
main part of the spectrum, the distribution has the shape of a
semicircle and not of a Gaussian. Such details should, however, be
irrelevant for a rough estimate). Under this assumption, the
fluctuation of the binding energy of a nucleus with $A$ nucleons
is obtained by the convolution of $(A - A_0)$ Gaussian distributions,
each with its own $A$--dependent width. The resulting distribution
is again Gaussian. Its variance $(\sigma_A)^2$ is the sum of the
variances of the individual Gaussians, the latter being given by the
$A$--dependent r.h.s. of Eq.~(\ref{4}). Thus,
\be
(\sigma_A)^2 = (30 \ {\rm keV})^2 + \sum_{A' = A_0 + 1}^A \frac{c}{N}
\biggl( \frac{\Gamma^{\downarrow} \lambda}{2 E_0} \biggr)^2 \ .
\label{7}
\ee
Eq.~(\ref{7}) holds provided that the fluctuations in different
nuclei are statistically uncorrelated. We are not aware of any
evidence to the contrary. The most significant dependence upon $A'$
on the r.h.s. of Eq.~(\ref{7}) resides in $N = \pi \lambda / d$.
Since $N$ increases strongly (nearly exponentially) with $A$, the
individual terms under the sum decrease rapidly, and $\sigma_A$ is
a monotonically increasing function of $A$ with ever decreasing
slope. This statement applies in general, quite irrespective of any
details of the $A$--dependence of $N$. We roughly model the
$A$--dependence of the terms under the sum by writing
\be
\frac{c}{N} \biggl( \frac{\Gamma^{\downarrow} \lambda}{2 E_0}
\biggr)^2 \approx (30 \ {\rm keV})^2 \exp ( \sqrt{A_0} - \sqrt{A} )
\ .
\label{8}
\ee
This formula takes account of the initial condition at $A_0 = 26$ and
models the $A$--dependence in terms of the exponential which roughly
reflects the dependence of the average level density on $A$. (A more
precise expression would have $\exp \sqrt{A}$ replaced by $\exp [
\alpha \sqrt{A}]$, with $\alpha$ of order unity and weakly dependent
on $A$. We disregard such corrections in view of the rather rough
overall approach we have taken). We use this expression in
Eq.~(\ref{7}), replace the summation by an integration, and obtain
\be
\sigma(A) = (30 \ {\rm keV}) \biggl[ 1 + 2 (\sqrt{A_0} + 1) (1 -
\frac{\sqrt{A_0} + 1}{\sqrt{A} + 1} \exp \{ \sqrt{A_0} - \sqrt{A} \}
) \biggr]^{1/2} \ .
\label{9}
\ee

\begin{figure}
\begin{center}
\includegraphics[clip,height=5cm]{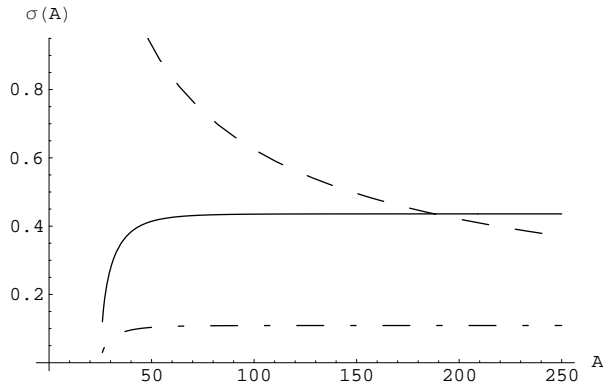}
\caption{\label{fig2}\textit{Fluctuations of nuclear
    ground--state energies. Dashed curve: fit to the experimental data
    taken from Ref.~\cite{Moe}; dot--dashed curve: formula (9) of the
    text; continuous curve: formula (9) with the factor 30 replaced by
    120 $keV$.}
 }
\end{center}
\end{figure}

In Figure~\ref{fig2}, we compare our Eq.~(\ref{9}) with the result
reported in Ref.~\cite{boh} for the fluctuations of the binding
energies versus $A$. To emphasize the uncertainties in our estimates,
we show two curves, one of which represents Eq.~(\ref{9}) as it stands
while for the other, we have arbitrarily replaced the factor 30 keV on
the r.h.s. of Eq.~(\ref{9}) by 120 keV. We see that for heavy nuclei,
the data and our result have similar values while for $A \approx 50$,
the difference is larger. In view of the rather large uncertainties of
our estimates, we would be satisfied with an order--of--magnitude
agreement and do not attach much significance to such statements. More
important, in our view, is the fact that we predict a monotonically
increasing function of $A$ while both, the data and the chaotic model
of Ref.~\cite{boh}, yield a monotonic decrease with $A$. As for the
data, we can think of additional effects that might affect the
$A$--dependence. For instance, it is conceivable that for light
nuclei, the clean separation between surface terms and volume terms
assumed in the mass formulas, does not apply. This might add a
correction which strongly decreases with $A$ and which has to be
removed from the data before a comparison with our results can be
made. Within the conservative phenomenological approach which we have
taken, there seems no way of escaping the conclusion that $\sigma_A$
increases monotonically with $A$. This is clear from Eq.~(\ref{7})
which, in turn, rests upon our identification of fluctuations of the
binding energy of a nucleus with mass number $A$ with the fluctuations
of the ground--state energies with respect to neutron or proton
threshold of all nuclei with mass numbers $A' < A$.

\end{document}